\documentclass[aps,pra,reprint,amsmath,amssymb,showpacs,nobibnotes,superscriptaddress]{revtex4-2}

\usepackage{graphicx}
\usepackage{bm}     
\usepackage{hyperref} 
\usepackage{xcolor}
\hypersetup{
    colorlinks,
    linkcolor={blue!80!black},
    citecolor={blue!80!black},
    urlcolor={blue!80!black}
}
\usepackage{dcolumn} 
\usepackage{dsfont}    
\usepackage{color}
\usepackage{textcomp,gensymb} 
\usepackage{verbatim} 
\usepackage{multirow} 
\usepackage{makecell} 
\usepackage{ulem} 

\makeatletter

\newif\ifequalcontrib
\newcommand{\equalcontribmark}{\global\equalcontribtrue}

\def\doauthor#1#2#3{%
  \global\equalcontribfalse
  \ignorespaces#1\unskip\@listcomma
  \begingroup
    #3%
  \@if@empty{#2}%
    {\endgroup{}{}}%
    {\endgroup{\comma@space}{}\frontmatter@footnote{#2}}%
  \ifequalcontrib
    \textsuperscript{,\ensuremath{\dagger}}%
  \fi
  \space\@listand
}
\makeatother

\begin{document}

\title{Scalability and performance metrics for boson sampling devices}

\author{Dong-Gil Im\equalcontribmark}
\email{eastgil7902@gmail.com}
\affiliation{Agency for Defense Development, Daejeon, 34186, Korea}

\author{Chung-Hyun Lee}
\thanks{These authors contributed equally to this work}
\affiliation{Department of Physics, Pohang University of Science and Technology (POSTECH), Pohang, 37673, Korea}

\author{Joonsuk Huh}
\email{joonsukhuh@gmail.com}
\affiliation{Department of Chemistry, Yonsei University, Seoul 03722, Korea}
\affiliation{Department of Quantum Information, Yonsei University, Incheon 21983, Korea}
\affiliation{Department of Computational Science and Engineering, Yonsei University, Seoul 03722, Korea}

\author{Yoon-Ho Kim}
\email{yoonho72@gmail.com}
\affiliation{Department of Physics, Pohang University of Science and Technology (POSTECH), Pohang, 37673, Korea}

\date{\today}

\begin{abstract}
Linear-optical platforms provide a promising route to demonstrate quantum computational advantage, particularly through the boson sampling problem. While a boson sampling system should be large enough to show the quantum advantage, the size of the boson sampling problem is often restricted by experimental constraints. In particular, as a boson sampling system is scaled up, experimental constraints can limit its effective problem size to permit efficient classical simulation. In this paper, we investigate the theoretical scalability bounds for boson sampling systems under practical experimental conditions, including photon loss, mode mismatch, and input state error. We employ non-negativity conditions for phase-space quasiprobability distribution functions to analyze these effects. 
In addition, we introduce two device-level metrics, the boson sampling score which measures the usable capacity of a given device and the boson sampling volume which provides a common hardware-level benchmark for comparing different platforms and assessing their access to an advantage-relevant problem size regime. Our results provide fundamental insights into the performance of boson sampling systems under realistic experimental conditions and offer practical guidelines for scaling up boson sampling systems.
\end{abstract}

\maketitle
\section{Introduction}

Demonstrating quantum computational advantage has become a central objective in quantum information science \cite{Boixo18}. In such a demonstration, a quantum device solves a well-defined task that is intractable for classical computers. This provides both a probe of computational complexity and a benchmark for comparing quantum platforms \cite{Harrow17,Zhong20}. Among the various approaches being explored for this purpose, boson sampling has emerged as a particularly promising candidate \cite{Gard14, Olson17}. In boson sampling, photons propagate through a linear-optical interferometer and are then measured by photon counting detectors. The resulting samples follow a probability distribution determined by the modulus squared of matrix permanents \cite{Valiant79, Aaronson13, Lund17}. For Gaussian boson sampling (GBS), squeezed states are injected instead of single photons and the output distribution is governed by hafnians \cite{Hamilton17}.
Since exactly evaluating matrix permanents and hafnians is \#P-hard \cite{Aaronson11}, large-scale boson sampling is believed to be intractable on a classical computer. Realizing quantum computational advantage with boson sampling requires genuinely nonclassical input states that cannot be expressed as mixtures of coherent states, since otherwise the sampling can be efficiently simulated on a classical computer \cite{Rahimi15}. Experiments have therefore employed single-photon Fock states or squeezed vacuum states as input states \cite{Broome13, Spring13, Wang19, Zhong21}.

The primary resource enabling quantum advantage in boson sampling is the number of indistinguishable particles contributing to quantum interference \cite{Dalzell20, Li21}. In practice, however, experimental constraints reduce the number of indistinguishable particles and thereby limit the effective size of the sampling problem \cite{Rohde12}. These constraints, such as loss and mode mismatch, accumulate as the system is scaled up. They make it challenging to realize boson sampling beyond the reach of classical computers \cite{Aaronson16, Rahimi16, Garcia19}. 
Since such a device may admit efficient classical simulation despite its nonclassical input states \cite{Renema18, Shchesnovich19}, quantifying how these constraints limit the scalable problem size of boson sampling in practice is essential.

In this paper, we investigate the relationship between the scalability of boson sampling and experimental constraints using non-negativity conditions of phase-space quasi-probability distribution (PQD) functions \cite{Rahimi16}. To determine the scalability bounds, we derive a sufficient condition for efficient classical simulation based on the non-negativity of the three generalized PQD functions: the input state PQD function, the channel transition function, and the measurement PQD function. These functions appear in the phase-space representation of the output probability. 
Based on this analysis, we introduce two device-level metrics: the boson sampling score and the boson sampling volume. The boson sampling score quantifies the usable capacity of a device by estimating the effective number of indistinguishable photons available for quantum interference at a given interferometer size. We then define the boson sampling volume by estimating the largest advantage-relevant problem size accessible to a given device, thereby providing a common hardware-level benchmark for comparing different platforms.
Although the volume does not certify quantum advantage by itself, it enables quantitative comparison of how efficiently different boson sampling devices preserve quantum interference relevant bosonic resources under experimental constraints.

We further apply the proposed metrics to a few recent Gaussian boson sampling experiments and assess their effective problem sizes. The analysis shows that increasing the number of detected photons does not necessarily increase the computationally relevant problem size. Once the number of optical modes becomes the main limitation, further scaling requires enlarging the number of optical modes while maintaining low channel loss and high photon indistinguishability.
Our results provide fundamental insights into the performance of boson sampling systems with experimental constraints and offer practical guidelines for scaling up boson sampling systems.

\section{PQD-based classical simulatability of boson sampling under experimental constraints}

\begin{figure}[t]
\centering
\includegraphics[width=0.85\columnwidth]{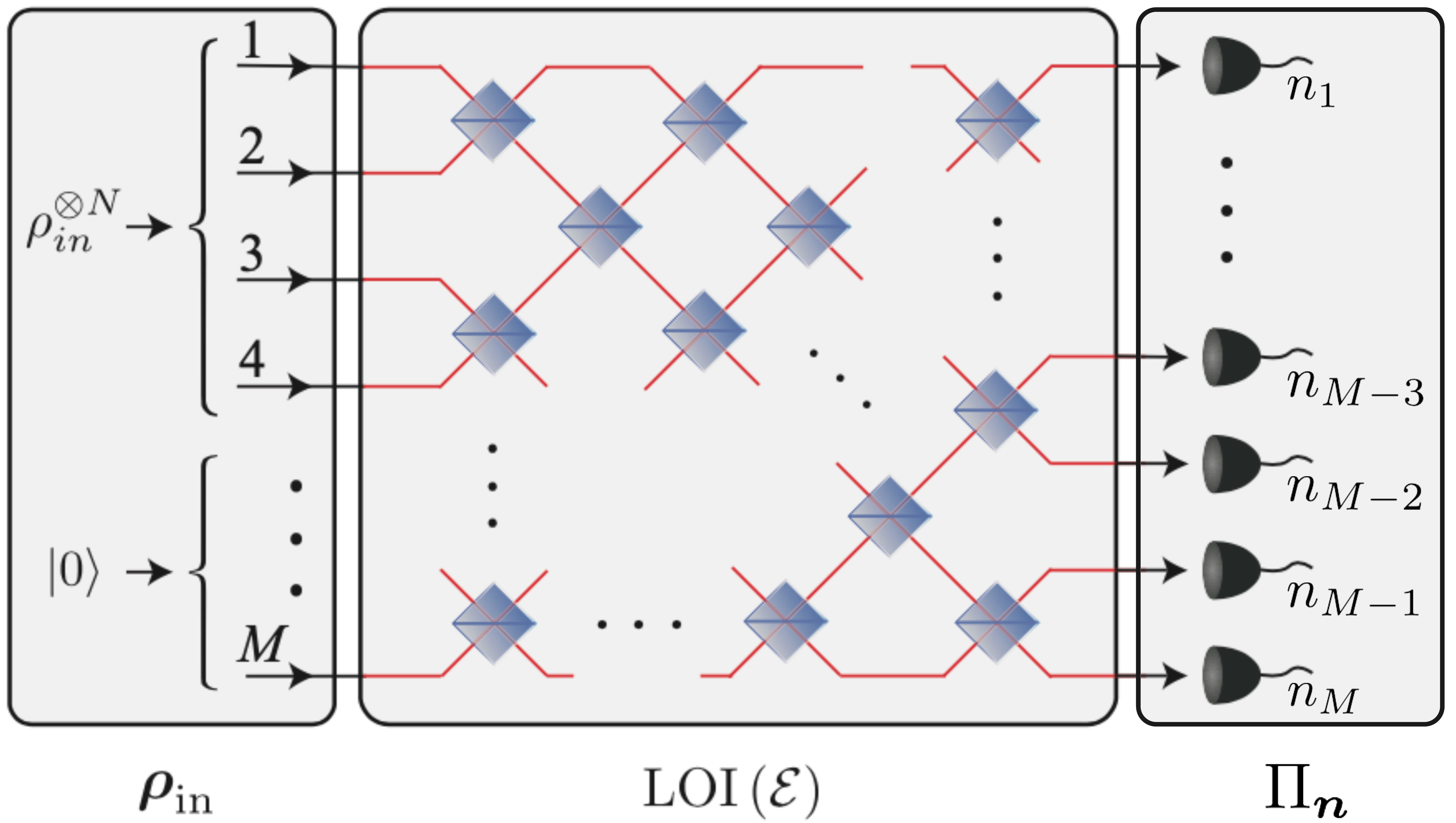}
\caption{Schematic of the quantum photonic device for boson sampling. The identical input states $\boldsymbol{\rho}_{\text{in}}$ are injected into the $N$  input modes of the $M$-port linear optical interferometer (LOI) described by the linear optical channel $\mathcal{E}$. After passing through the interferometer, the photons in the output modes are measured by the on-off photodetectors with the measurement operator $\boldsymbol{\Pi}_{\boldsymbol{n}}$, where ${\boldsymbol{n}}$ is the output photon number vector ${\boldsymbol{n}} = (n_1, n_2, \cdots, n_M)$. 
}\label{fig:BS}
\end{figure}

Consider the schematic of the boson sampling device represented in Fig.~\ref{fig:BS}. The device involves the injection of $N$ identical input states $\boldsymbol{\rho}_{\text{in}}$ into an $M$-port linear optical interferometer (LOI) described by a linear optical channel $\mathcal{E}$. The photons traverse the interferometer and are measured by photodetectors employing the measurement operator $\boldsymbol{\Pi}_{\boldsymbol{n}}$, where ${\boldsymbol{n}} = (n_1, n_2, \cdots, n_M)$ is the output photon number vector. In this work, we assume the use of on-off photodetectors ($n_{i} = 0$ or 1), as they align with the prevalent choice of single-photon detectors in current quantum optics experiments \cite{ Kim11, Natarajan12}. 
To calculate the measurement result of the boson sampling system, we start with the output probability $p(\boldsymbol{n}) = \text{Tr}[\mathcal{E}(\boldsymbol{\rho}_{\text{in}}) \boldsymbol{\Pi}_{\boldsymbol{n}}]$ expressed by the generalized phase-space quasiprobability distribution (PQD) functions represented in Ref.~\cite{Rahimi16},
\begin{equation}
\begin{split}\label{eq:Output}
p(\boldsymbol{n}) =& \,\pi^M \int d^{2M} \boldsymbol{\beta} \int d^{2M} \boldsymbol{\alpha} \, W^{(-\boldsymbol{s})}_{\boldsymbol{\Pi_{\boldsymbol{n}}}} (\boldsymbol{n}|\boldsymbol{\beta}) 
\, T_{\mathcal{E}}^{(\boldsymbol{s},\boldsymbol{t})} (\boldsymbol{\beta}|\boldsymbol{\alpha})  
\\ & \times W^{(\boldsymbol{t})} (\boldsymbol{\alpha}|\boldsymbol{\rho}_{\text{in}}),
\end{split}
\end{equation}
where $W^{(\boldsymbol{t})} (\boldsymbol{\alpha}|\boldsymbol{\rho}_{\text{in}})$, $T_{\mathcal{E}}^{(\boldsymbol{s},\boldsymbol{t})} (\boldsymbol{\beta}|\boldsymbol{\alpha})$, and $W^{(-\boldsymbol{s})}_{\boldsymbol{\Pi_{\boldsymbol{n}}}} (\boldsymbol{n}|\boldsymbol{\beta})$ are the generalized PQD function for the input state $\boldsymbol{\rho}_{\text{in}}$, the quasi-probability transition function for the linear optical channel $\mathcal{E}$, and the generalized PQD function for the output measurement $\boldsymbol{\Pi}_{\boldsymbol{n}}$, respectively. The generalized PQD functions are parameterized by the diagonal ordering matrices $\boldsymbol{s} = \text{diag}(s_1, s_2, \cdots, s_M)$ and $\boldsymbol{t} = \text{diag}(t_1, t_2, \cdots, t_M)$ where each diagonal element specifies the ordering parameter for the corresponding optical mode. The explicit forms of generalized PQD functions are introduced in the following subsection. 
Here, we denote the input and output phase-space amplitudes by the complex vectors $\boldsymbol{\alpha} = (\alpha_1, \alpha_2, \cdots, \alpha_M)$ and $\boldsymbol{\beta} = (\beta_1, \beta_2, \cdots, \beta_M)$, respectively. (Throughout, for vectors of complex numbers, e.g., $\boldsymbol{\gamma} = \boldsymbol{\alpha}$, $\boldsymbol{\beta}$, $\hat{a}_j |\boldsymbol{\gamma}\rangle = \gamma_j |\boldsymbol{\gamma}\rangle$ is satisfied, where $\hat{a}_j $ is the annihilation operator for the $j$th mode and $|\boldsymbol{\gamma} \rangle = \bigotimes^M_{j=1} |\gamma_j \rangle$ is the multimode coherent state.)

To identify the scalability bounds of boson sampling, it is necessary to establish a sufficient condition under which boson sampling devices become classically simulatable. If all three PQD functions are non-negative and can be efficiently sampled with classical computational resources that scale polynomially with $M$, the output probability distribution $p(\boldsymbol{n})$ can be sampled by the following sequential procedure: For a given input state $\boldsymbol{\rho}_{\text{in}}$, choose the vector of complex amplitudes $\boldsymbol{\alpha}$ from $W^{(\boldsymbol{t})} (\boldsymbol{\alpha}|\boldsymbol{\rho}_{\text{in}})$. Then, for the given $\boldsymbol{\alpha}$, choose the vector of complex amplitudes $\boldsymbol{\beta}$ from  $T_{\mathcal{E}}^{(\boldsymbol{s},\boldsymbol{t})} (\boldsymbol{\beta}|\boldsymbol{\alpha})$. Finally, for the given $\boldsymbol{\beta}$, choose $\boldsymbol{n}$ from $W^{(-\boldsymbol{s})}_{\boldsymbol{\Pi_{\boldsymbol{n}}}} (\boldsymbol{n}|\boldsymbol{\beta})$. Thus, the non-negativity of the three PQD functions provides a sufficient condition for efficient classical simulation \cite{Rahimi16}.  

Based on the phase-space formulation above, we now derive the non-negativity conditions for each of the three PQD functions and combine them into a single criterion applicable to realistic boson sampling implementations.

\subsection{The non-negativity conditions of boson sampling}


First, we determine the non-negativity condition for the $(\boldsymbol{t})$-ordered generalized PQD function of the input state $\boldsymbol{\rho}_{\text{in}}$, which is defined as
\begin{equation}
\begin{split}\label{eq:inputPQD}
W^{(\boldsymbol{t})} (\boldsymbol{\alpha}|\boldsymbol{\rho}_{\text{in}}) = \int \frac{d^{2M} \boldsymbol{\xi}}{\pi^{2M}} \Phi^{(\boldsymbol{t})} (\boldsymbol{\xi}|\boldsymbol{\rho}_{\text{in}}) 
e^{\boldsymbol{\alpha} \boldsymbol{\xi}^\dagger - \boldsymbol{\xi} \boldsymbol{\alpha}^\dagger} ,
\end{split}
\end{equation}
where $\Phi^{(\boldsymbol{t})} (\boldsymbol{\xi}|\boldsymbol{\rho}_{\text{in}}) = \text{Tr}[\boldsymbol{\rho}_{\text{in}} D(\boldsymbol{\xi})] e^{\boldsymbol{\xi} \boldsymbol{t} \boldsymbol{\xi}^\dagger/2} $ with $M$-mode displacement operator $D(\boldsymbol{\xi}) = e^{\boldsymbol{\xi} \boldsymbol{a}^\dagger - \boldsymbol{a} \boldsymbol{\xi}^\dagger }$, $\boldsymbol{a} = (a_1,\cdots, a_M)$ the row vector of modal annihilation operators, and $\boldsymbol{a}^\dagger = (a_1^\dagger,\cdots, a_M^\dagger)^T$ the column vector of creation operators  \cite{Cahill69,Hillery84}. Here, $\boldsymbol{\xi}$ is the vector of complex amplitudes, $\boldsymbol{\xi} = (\xi_1, \xi_2, \cdots, \xi_M)$.
Specifically, the ($\boldsymbol{t}$)-ordered generalized PQD function gives the $M$-mode Husimi $Q$ function \cite{Cartwright76}, the Wigner function \cite{Kenfack04}, and the Glauber-Sudarshan $P$ function \cite{Glauber63,Kiesel08} for $\boldsymbol{t} = -\boldsymbol{I}_{M}$, $\boldsymbol{t} = \boldsymbol{O}_M$, and $\boldsymbol{t} = \boldsymbol{I}_{M}$, respectively. Here, $\boldsymbol{I}_{M}$ and $\boldsymbol{O}_M$ are the $M$-dimensional identity matrix and the $M$-dimensional zero matrix, respectively. 

For the single-mode input state $\rho_{\text{in}}$, we define $\bar{t}$ as the largest scalar ordering parameter for which its PQD function is non-negative. Then, the non-negativity persists for every $t \leq \bar{t}$. With the first $N$ modes populated, the $M$-mode input state is $\boldsymbol{\rho}_{\text{in}} = \rho_{\text{in}}^{\otimes N} \otimes (|0\rangle \langle 0|)^{\otimes (M-N)}$ and the multimode input PQD is non-negative when 
\begin{equation}
\begin{split}\label{eq:inputnonnegative}
\boldsymbol{t} \leq  \bar{t} \boldsymbol{J}_{N} + (\boldsymbol{I}_{M}-\boldsymbol{J}_{N}),
\end{split}
\end{equation}
where $\boldsymbol{J}_{N} = \operatorname{diag} \left( \underbrace{1,\ldots,1}_{N}, \underbrace{0,\ldots,0}_{M-N} \right)$. Thus, $t_i \leq \bar{t}$ in a populated mode, whereas $t_i \leq 1$ is allowed in a vacuum mode since the vacuum state has a non-negative Glauber-Sudarshan $P$ representation. 

Second, we find the non-negativity condition for the quasi-probability transition function $T_{\mathcal{E}}^{(\boldsymbol{s},\boldsymbol{t})} (\boldsymbol{\beta}|\boldsymbol{\alpha})$ associated with the linear optical channel $\mathcal{E}$. Consider the case where the channel is a lossy $M$-port LOI. Then, the quasi-probability transition function is given by \cite{Rahimi16},
\begin{equation}
\begin{split}\label{eq:channel}
T_{\mathcal{E}}^{(\boldsymbol{s},\boldsymbol{t})} (\boldsymbol{\beta}|\boldsymbol{\alpha}) \propto \exp  [ -2 ( \boldsymbol{\beta}-\boldsymbol{\alpha} \boldsymbol{L} ) \boldsymbol{\Sigma}^{-1} ( \boldsymbol{\beta}^{\dagger}-\boldsymbol{L}^{\dagger} \boldsymbol{\alpha}^{\dagger} )  ] \big / \det \boldsymbol{\Sigma},
\end{split}
\end{equation}
where $\boldsymbol{L}$ is the $M\,\times \, M$ matrix describing the LOI and $\boldsymbol{\Sigma} = \boldsymbol{I}_{M}-\boldsymbol{L}^{\dagger}\boldsymbol{L}-\boldsymbol{s}+ \boldsymbol{L}^{\dagger} \boldsymbol{t} \boldsymbol{L}$. 
When $\boldsymbol{\Sigma} \geq 0$, the quasi-probability transition function is non-negative and can be interpreted as a conditional probability distribution for sampling the output phase-space amplitude $\boldsymbol{\beta}$ given the input amplitude $\boldsymbol{\alpha}$. Note that this condition concerns only the non-negativity of the channel representation for the chosen ordering parameters and does not characterize the interferometer itself as classical or nonclassical.
 For simplicity, we assume that each input photon experiences the same amount of loss with a constant transmission $\eta_{L} $ and thus describe the interferometer by a matrix $\boldsymbol{L} = \sqrt{\eta_L} \boldsymbol{U}$ where $\boldsymbol{U}$ is a unitary matrix for a lossless channel. Hence, the non-negativity condition for the quasi-probability transition function of the linear optical channel is given by
\begin{equation}
\begin{split}\label{eq:channelNegat}
 \boldsymbol{\Sigma} =  (1-\eta_{L})\boldsymbol{I}_{M}-\boldsymbol{s}+\eta_{L}\boldsymbol{U}^{\dagger}\boldsymbol{t}\boldsymbol{U} \geq 0.
\end{split}
\end{equation}

Third, the non-negativity condition for the measurement PQD function $W^{(-\boldsymbol{s})}_{\boldsymbol{\Pi_{\boldsymbol{n}}}} (\boldsymbol{n} | \boldsymbol{\beta})$ for on-off detection is found as \cite{Rahimi16}
 \begin{equation}
\begin{split}\label{eq:Measure}
\boldsymbol{s} \geq \boldsymbol{I}_{M}-(2/\eta_D)\boldsymbol{p}_D,
\end{split}
\end{equation}
 where $\eta_D$ is the detector efficiency, and $\boldsymbol{p}_D$ is the diagonal matrix of random count probability of each detector. Here, we assume that the on-off photodetectors assigned to all output modes have the same detection efficiency and dark count probability for simplicity.

Finally, combining the non-negativity conditions of PQD functions for the input state, linear optical channel, and output measurement yields the simplified condition, which can be represented by
\begin{equation}
\begin{split}\label{eq:singleCond}
 \boldsymbol{\Sigma}  = \left(2/{\eta_D}\right)  \boldsymbol{p}_D + \eta_L (\bar{t}-1) \, \boldsymbol{U}^{\dagger} \boldsymbol{J}_{N} \boldsymbol{U} \geq 0.
\end{split}
\end{equation}
Once $\eta_{L}$ and $\boldsymbol{p}_{D}$ are specified by the adopted loss and noise models, respectively, Eq.~(\ref{eq:singleCond}) provides a sufficient condition for efficient classical sampling from the boson-sampling output distribution. Specifically, if Eq.~(\ref{eq:singleCond}) is satisfied, the output distribution can be sampled efficiently using the sequential phase-space procedure described above. 

\subsection{Analysis of mode mismatch}
In the context of implementing boson sampling, one of the primary errors in the system is mode mismatch \cite{Rahimi16, Rohde12,Aaronson16, Garcia19}. This phenomenon occurs when, for instance, the temporal, frequency, and polarization properties of photon wave packets do not overlap perfectly. As a consequence, the non-overlapping portions of photon wave packets do not contribute to the quantum interference necessary for generating the desired output. However, these components can still propagate through the interferometer and produce detection events when they fall within the spatio-temporal windows of the single-photon detectors. This effect becomes particularly critical when mode mismatching dominates the random count probability, especially when the intrinsic dark-count rate of high-quality detectors is exceptionally low. 

To incorporate mode mismatch into the PQD-based simulatability condition, we introduce the indistinguishability parameter $p_{ind}$, with $0 \leq p_{ind} \leq 1$. We interpret $p_{ind}$ as the fraction of the indistinguishable photons, which preserves the intended quantum interference. By contrast, the complementary fraction $1-p_{ind}$ occupies distinguishable internal modes and does not contribute coherently to the interference process. We account for these two components separately in the channel and detection models. Since only the indistinguishable fraction contributes to the quantum interference, the effective transmission in the second term of Eq.~(\ref{eq:singleCond}) is reduced from $\eta_L$ to $\eta_L p_{ind}$. The distinguishable fraction $1-p_{ind}$ may nevertheless propagate through the interferometer and generate detector clicks, which we treat as an incoherent background contribution.

Since only $N$ of the $M$ input modes are populated, the mode mismatch induced background generally depends on the fraction of occupied input ports in the $M$-mode interferometer. For a sufficiently mode-mixing interferometer, the distinguishable contribution is approximately distributed uniformly over the $M$ outputs. Defining the filling fraction as $\nu = N / M$, the mode mismatch induced random count matrix is given as $\boldsymbol{p}_{mis} \simeq \mu \eta_L \eta_D (1-p_{ind}) \nu \boldsymbol{I}_M$, where $\mu$ denotes the mean photon number per populated input port.
Including the intrinsic dark-count probability $D_c$, the total random count matrix entering the detector condition becomes $\boldsymbol{p}_{D} = \boldsymbol{p}_{mis} + D_{c} \boldsymbol{I}_M$. Substituting this expression into Eq.~(\ref{eq:singleCond}), together with the reduced interference-preserving transmission $\eta_L p_{ind}$, gives
\begin{equation}
\label{eq:threshold}
{2 \mu\eta_{L}(1-p_{ind})} \nu+ \frac{2D_{c}}{\eta_{D}}+\eta_{L}p_{ind} (\bar{t}-1) \geq 0.
\end{equation}
The general interferometer dependent form of the mode mismatch induced background and the derivation of the uniform-mixing approximation are provided in Sec.~I of the Supplementary Material. 

Under the equal-overlap model adopted here, where all input photons are assumed to have the same pairwise wave-packet fidelity $\mathcal{F}$, the indistinguishability parameter is related to $\mathcal{F}$ by $p_{ind} = \sqrt{\mathcal{F}}$. The corresponding internal-mode decomposition is given in Sec.~I of the Supplementary Material.

\section{Scalability Bounds of Boson Sampling and performance estimation}

In this section, we explore the scalability bound of the boson sampling problem in real devices using Eq.~(\ref{eq:threshold}). 
If $2\mu(1-p_{ind}) \nu + p_{ind}(\bar{t}-1) \geq 0$, Eq.~(\ref{eq:threshold}) is trivially satisfied for any physical transmission $0 \leq \eta_L \leq 1$, since both $\eta_L$ and $2D_c/{\eta_D}$ are non-negative. In this case, the sufficient condition for efficient classical simulation is already fulfilled. Therefore, the corresponding boson sampling device is classified as classically simulatable within the present PQD-based sufficient criterion. We therefore focus on the nontrivial case, $2\mu(1-p_{ind}) \nu + p_{ind}(\bar{t}-1)<0$. We further assume that the transmission of the linear optical interferometer is a monotonically decreasing continuous function $\eta_L(M)$ with respect to the number of interferometer ports $M$. Then, we can rewrite Eq.~(\ref{eq:threshold}) as
\begin{equation}
\label{eq:rewrite eq.2}
\eta_{L}(M) \leq \eta_{th},
\end{equation}
where $\eta_{th} = - 2D_{c}/(2\mu \eta_{D} (1-p_{ind}) \nu+ \eta_{D} p_{ind} \left ( \bar{t} -1 \right )) $.
Then, we can determine the threshold of $M$ that makes the boson sampling device classically simulatable, as given by
\begin{equation}
\label{eq:Scalability_Bound}
M \geq \eta_{L}^{-1} \left ( \eta_{th} \right ),
\end{equation}
where $\eta_{L}^{-1}$ is an inverse function of $\eta_{L}$. Here, we assume that the input filling fraction $\nu = N / M$ remains fixed as the interferometer is scaled. Since the performance of boson sampling is constrained by the number of modes of the interferometer, we identify 
\begin{equation}
\label{eq:M_B}
M_{B} = \left \lceil \eta_{L}^{-1} (\eta_{th}) \right \rceil
\end{equation}
 as a scalability bound of the boson sampling experiment.

\subsection{Boson sampling score $S_{BS}$ and volume $V_{BS}$}
We identify the scalability bound of the boson sampling device, yet this bound does not directly indicate its performance. The performance of a boson sampling device depends on the size of the submatrix used for calculating the matrix permanent or hafnian \cite{Aaronson13, Hamilton17}. If the size of the submatrix is small, the boson sampling computing power can be worse than that of a classical computer, even with a large number of interferometer ports \cite{Aaronson16, Rohde12}.

There are three key parameters that determine the performance of a boson sampling device: the number of injection ports $N$; the number of ports of the interferometer $M$; and the total number of detected photons across all output ports, $N_D$. In the case of boson sampling with single-photon Fock states, $N$ plays a pivotal role in determining the size of the submatrix within the unitary matrix that describes the interferometer. For Gaussian boson sampling, $N$ represents the submatrix rank. Additionally, the total detected photon number $N_D$ is another crucial parameter that influences the size of the submatrix in both cases \cite{Rahimi16}.

To evaluate the usable capacity for a given device comprehensively, we introduce a single numerical metric called the boson sampling score, defined as 
\begin{equation}
\label{eq:Score}
S_{BS} = \begin{cases} 
\min [ N_{sat}, M ] \times \min [ \mu p_{ind} \eta_{L} \eta_{D}, 1]
&  M < M_{B} \\
\\
0
& M \geq M_{B}
\end{cases}
\end{equation}
where $N_{sat} = M / (\mu \eta_{L} \eta_{D})$ is the number of populated input ports at which the expected total detected photon number reaches $M$. Accordingly, the term $ \min[ N_{sat}, M]$ in Eq.~(\ref{eq:Score}) represents the effective number of input mode slots that can be populated before the mean detected occupancy reaches one photon per output mode. The second term $\min [\mu p_{ind} \eta_{L} \eta_{D}, 1 ]$ denotes the mean number of detected photons per populated input port that remain indistinguishable and can participate in interference. Therefore, the boson sampling score $S_{BS}$ reflects the effective number of detected, indistinguishable photons available for quantum interference for a given device. This, in turn, serves as a proxy for the effective matrix size relevant to the permanent or hafnian calculation. We set $S_{BS} = 0$ when $M \geq M_B$, since the PQD-based criterion then admits an efficient classical simulation. 

Next, we define the boson sampling volume $V_{BS}$ as a common hardware-level benchmark for assessing a device's access to an advantage-relevant problem size regime:
\begin{equation}
\label{eq:Volume}
V_{BS}(M_{L}) = \max_{M \leq M_{L}} \left [ \min \left [ S_{BS} (M), \sqrt{M} \right ] \right ],
\end{equation}
where $M_{L}$ represents the practical constraint on the mode size of the interferometer. 
Here, the factor $S_{BS}(M)$ estimates the average number of such effective detected photons for a given interferometer size $M$. The factor $\sqrt{M}$ sets the characteristic scale for the low-collision regime, where a collision refers to two or more photons occupying the same output mode. Since the expected number of collisions scales approximately as $N_D^2/M$, collisions remain rare when $N_D$ is below the order of $\sqrt{M}$. We therefore use $N_D \lesssim \sqrt{M}$ as the low-collision constraint in defining $V_{BS}$. The maximization over $M$ then identifies the largest effective problem size accessible to a given experimental platform before either experimental limitations or collision effects become dominant.
Thus, the boson sampling volume $V_{BS}$ has a direct operational interpretation as the largest effective low-collision boson sampling problem size that a realistic device can support after accounting for input brightness, photon loss, detector efficiency, mode mismatch, and the classical simulability bound imposed by the non-negativity of PQD functions.
A systematic procedure for evaluating the maximization in Eq.~(\ref{eq:Volume}) for different transmission scalings is provided in Sec.~II of the Supplementary Material.

Importantly, $V_{BS}$ should not be interpreted as a proof of quantum advantage. 
Rather, it quantifies the effective low-collision problem size accessible to a given device. A small $V_{BS}$ indicates limited access to large low-collision instances. Conversely, a large value of $V_{BS}$ indicates that the device can support an equivalent ideal low-collision boson sampling problem size comparable to the regime believed to be classically hard, although additional requirements, such as validation tests \cite{Spagnolo14,Seron24} and comparison with optimized classical algorithms \cite{Oh24}, must still be addressed.

Together, $M_B$ and $V_{BS}$ provide a two-sided characterization of device performance. The bound $M_B$ identifies, within the present PQD model, a region in which efficient classical simulation is certified by the sequential phase-space sampling construction, without relying on benchmarking against a particular optimized classical algorithm. In contrast, the volume $V_{BS}$ quantifies the largest effective problem size within the low-collision regime $N_D \lesssim \sqrt{M}$, where standard hardness arguments for boson sampling are most directly applicable \cite{Aaronson13}. 
Outside both the PQD simulable and low-collision regimes, neither criterion provides a definitive assessment of quantum advantage. Claims of advantage must therefore be established by direct comparison with the best available classical algorithms. Thus, these quantities locate a device with respect to a certified-simulable boundary and an advantage-relevant low-collision boundary.

\subsection{Architecture-level requirements for target $V_{BS}$}
Building on the preceding discussion, $V_{BS}$ can also serve as a practical design metric for comparing different interferometer architectures. In particular, $V_{BS}$ connects component-level constraints, such as the transmission of individual optical elements, to the effective boson sampling problem size that can be realized experimentally. To illustrate this point, we estimate the component-level transmission required to achieve a given target $V_{BS}$ for three representative interferometer architectures: the universal architecture \cite{Flamini17,Crespi13}, the fast Fourier transform (FFT) architecture \cite{Takiguchi09, Crespi16}, and the Jiuzhang-type architecture \cite{Wang19, Zhong21, Zhong20}.

First, we model the transmission $\eta_L$ of interferometer $a$ as
\begin{equation}
\label{eq:InterfTrans}
\eta_L(M) = \eta_c \eta_0^{d_a(M)}.
\end{equation}
Here, $\eta_c$ is a coupling efficiency of the photon source to the LOI, $\eta_0$ is the effective transmission of each elementary optical unit, and $d_a(M)$ is the interferometer depth to fully connect $M$ number of modes. 

\begin{figure}[th!]
\centering
\includegraphics[width=0.84\columnwidth]{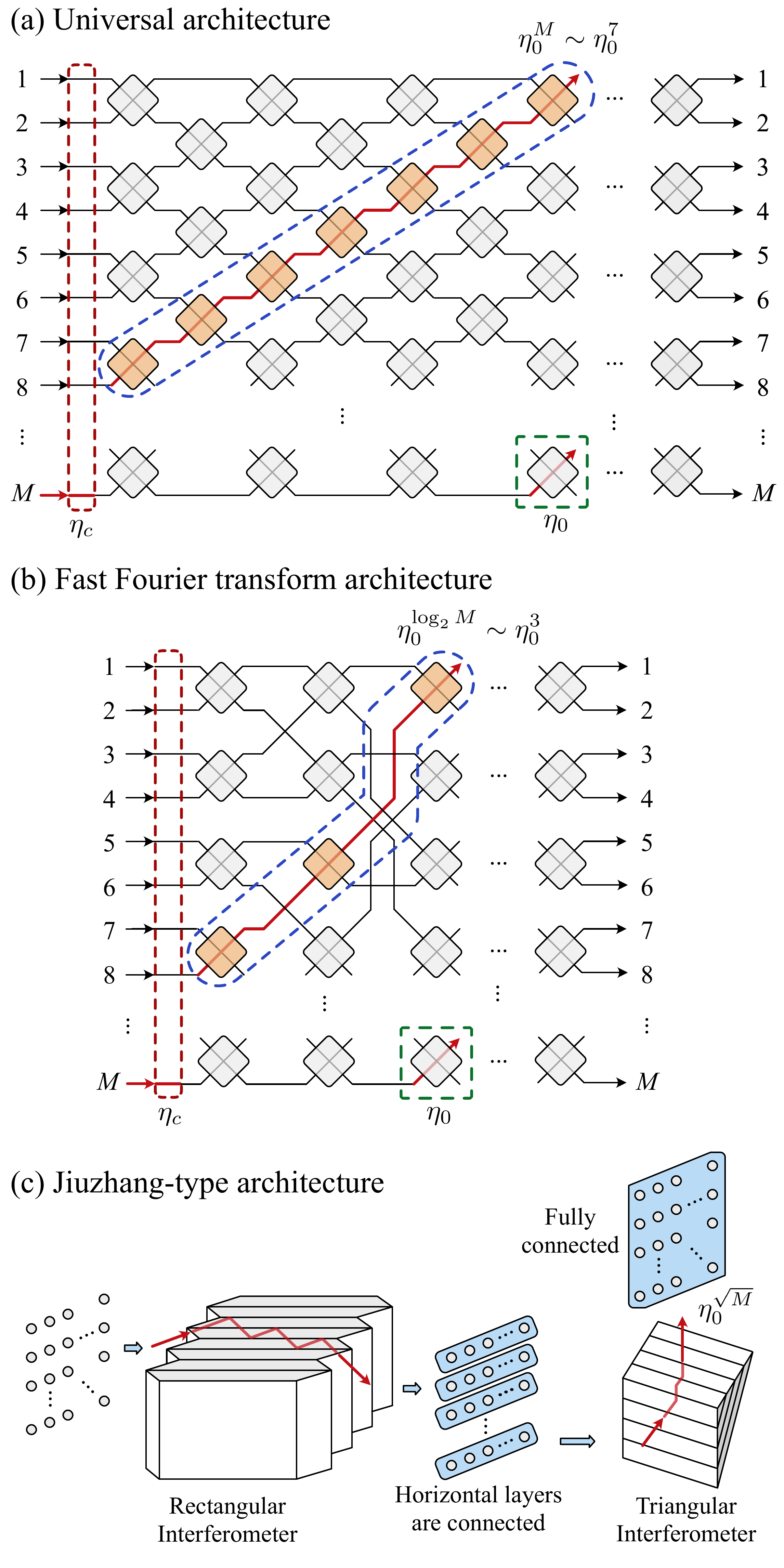}
\caption{Representative interferometer architectures for boson sampling. (a) Universal architecture. The blue dashed contour highlights a representative propagation route, and the orange elements indicate the optical units encountered along the route. The highlighted path gives a transmission of $\eta_0^7$ for the case of $M=8$, consistent with the architecture-level scaling $d_{uni}(M) \sim M$. The red dashed box at the input denotes the coupling efficiency $\eta_c$ of the photon source to the LOI, and the green dashed box identifies a single unit with transmission $\eta_0$. 
(b) Fast Fourier transform architecture. The highlighted path indicated by the blue dashed contour passes through three sequential optical units, giving a transmission of $\eta_0^3$ for the case of $M=8$, consistent with the architecture-level scaling $d_{fast}(M) = \log_2 M$.
(c) Jiuzhang-type architecture. The rectangular interferometer first connects modes within each horizontal layer. The subsequent triangular interferometer couples different layers, resulting in full connectivity among all modes. The red lines indicate a propagation route through the two interferometer blocks and the effective interferometer depth scales as $d_{Jiuzhang}(M) = \sqrt{M}$.
}\label{fig:archit}
\end{figure}

Figure \ref{fig:archit} shows the interferometers for three architectures. A universal architecture shown in Fig.~\ref{fig:archit}(a) realizes an arbitrary $M\times M$ unitary using a network of beam splitters and phase shifters with $\mathcal{O}(M)$ sequential depth \cite{Flamini17,Crespi13}. Therefore, we use $d_{uni}(M) = M$. 
The FFT architecture \cite{Takiguchi09, Crespi16}, inspired by the Cooley-Tukey fast Fourier transform algorithm, reduces the interferometer depth to a logarithmic scaling with the number of modes, $d_{fast}(M) = \log_2 M$ as shown in Fig.~\ref{fig:archit}(b). As the number of sequential optical units grows only logarithmically with $M$, this architecture can preserve a substantially larger effective problem size from the viewpoint of $V_{BS}$. 
The Jiuzhang-type architecture \cite{Wang19, Zhong21, Zhong20} employs a compact three-dimensional optical network designed to combine high transmission, high wave-packet overlap, and large-scale mode connectivity. As shown in Fig.~\ref{fig:archit}(c), the rectangular interferometer mixes modes within each horizontal layer, and the subsequent triangular interferometer couples different layers. Its effective interferometer depth scales approximately as $d_{Jiuzhang}(M) = \sqrt{M}$. This scaling places the Jiuzhang-type architecture between the universal and FFT architectures: it is substantially more tolerant to loss than a universal linear-depth interferometer, while still requiring higher component transmission than a logarithmic-depth architecture. The Jiuzhang-type architecture and the origin of this transmission scaling are described in Sec.~III of the Supplementary Material.

Using these transmission models, we estimate the effective transmission parameter $\eta_0$ required for each architecture to reach $V_{BS}\simeq 50$ and $V_{BS}\simeq 100$, as summarized in Table \ref{table:Volume}. The reference scale of 50 is motivated by the classical benchmarking of Gaussian boson sampling with threshold detection on the Sunway TaihuLight supercomputer \cite{Li21}. We adopt this scale as a design target, rather than as a threshold for classical intractability. The larger target of 100 is chosen to examine how the transmission requirements change when the effective low-collision problem size is doubled. Under our volume definition, this doubling increases the minimum required number of modes from 2,500 to 10,000, providing a concrete comparison of the loss accumulation in different architectures as the system is scaled up. We apply the same target volumes to both single-photon Fock-state and squeezed-vacuum inputs to compare their hardware requirements at a common effective problem size.

\begin{table}[t]
 \caption{Required effective transmission parameters for reaching target boson sampling volumes in three interferometer architectures. Here, we assume full input occupation, $N=M$, corresponding to $\nu=1$ and also relax the mode-number limitation of the interferometer by setting $M_L = \infty$.
Here, $\eta_c$ is the coupling efficiency of the photon source to the LOI, and $\eta_0$ is the effective transmission of each elementary optical unit. $\hat{S}(z) = \exp(\frac{1}{2} (z^* \hat{a}^2 - z \hat{a}^{\dagger \,2})) $ is a squeezing operator where $z = r e^{i \theta}$.
 }
 \begin{center}
\begin{tabular}{ccccc}
\noalign{\smallskip} \noalign{\smallskip}
\hline \hline
\noalign{\smallskip}
\multicolumn{2}{c}
{}        
& \begin{tabular}[c]{@{}c@{}}Universal\\ Architecture\\ \cite{Flamini17,Crespi13} \end{tabular} 
& \begin{tabular}[c]{@{}c@{}}FFT\\ Architecture\\ \cite{Takiguchi09, Crespi16} \end{tabular} 
& \begin{tabular}[c]{@{}c@{}}Jiuzhang\\ Architecture \\ \cite{Wang19, Zhong21, Zhong20} \end{tabular} \\ \hline \noalign{\smallskip}
\multicolumn{2}{c}
{$\eta_{L}$} 
& $\eta_{c} \eta_{0}^{M}$ 
& $\eta_{c} \eta_{0}^{\log_{2} M}  $  
& $\eta_{c} \eta_{0}^{\sqrt{M}}$

     \\ \hline \noalign{\smallskip}
\multirow{2}{*}{ \begin{tabular}[c]{@{}c@{}} $\eta_{0}$ for \\ $ V_{BS} (\infty) \simeq 50$\end{tabular} }    
& $\vert 1 \rangle$   &        0.9986 & 0.5999    &    0.9306
   \\ \cline{2-5}  \noalign{\smallskip}
& $\hat{S}(z) \vert 0 \rangle$   & 0.9980           &   0.5809      &   0.9047       
  \\ \hline \noalign{\smallskip}
\multirow{2}{*}{ \begin{tabular}[c]{@{}c@{}} $\eta_{0}$ for \\ $ V_{BS} (\infty) \simeq 100$\end{tabular}}   
& $\vert 1 \rangle$   & 0.9996        &          0.6127          &       0.9580
      \\ \cline{2-5}  \noalign{\smallskip}
   & $\hat{S}(z) \vert 0 \rangle $   & 0.9995          &  0.5971       &    0.9446   \\ \hline \hline
\end{tabular}
\end{center}
\label{table:Volume}
\end{table}

For the single-photon Fock-state boson-sampling case, we use experimentally motivated parameters $\eta_c=0.83$, mutual fidelity $\mathcal{F}=0.92$, mean photon number per input mode $\mu=1$, and $\bar{t}=-1$ \cite{Wang19}. For Gaussian boson sampling with squeezed vacuum states, we use $\eta_c=0.92$, $\mathcal{F}=0.96$, squeezing parameter $r=1.4$, $\mu=\sinh^2 r$, and $\bar{t}=e^{-2r}$ \cite{Zhong21}. For both cases, we take the detector efficiency and dark-count probability to be $\eta_D=0.92$ and $D_c=5\times10^{-6}$, respectively \cite{Zhong21}. Here, we relax the mode-number limitation of the interferometer by setting $M_L = \infty$ for simplicity.

The comparison in Table \ref{table:Volume} reveals a clear dependence of the required transmission on the interferometer depth scaling of each architecture. The universal architecture, whose interferometer depth scales linearly with $M$, requires nearly unity transmission, with $\eta_0=0.9986$ and 0.9980 for single-photon Fock and squeezed-vacuum inputs, respectively, to reach $V_{BS}\simeq50$, and values of approximately 0.9995 or higher to reach $V_{BS}\simeq100$. In contrast, the logarithmic-depth FFT architecture substantially relaxes this requirement, requiring $\eta_0\approx0.60$ and 0.61 for $V_{BS}\simeq50$ and 100, respectively, in the single-photon case, with slightly lower values for squeezed-vacuum inputs. The Jiuzhang-type architecture, with an intermediate interferometer depth scaling proportional to $\sqrt{M}$, correspondingly requires intermediate transmission values, $\eta_0=0.9306$ and 0.9580 for single-photon inputs and 0.9047 and 0.9446 for squeezed-vacuum inputs at $V_{BS}\simeq50$ and 100, respectively. 

Overall, Table \ref{table:Volume} demonstrates how the boson sampling volume can be used as a practical design criterion. Rather than simply increasing the nominal number of modes or input photons, one should optimize the architecture to maximize the effective problem size that remains accessible after all relevant constraints are taken into account. In this sense, $V_{BS}$ provides a bridge between experimental hardware parameters and the computationally relevant scale of the boson-sampling problem.

\subsection{Application to Gaussian boson sampling experiments}
It is useful to evaluate the proposed boson sampling score $S_{BS}$ and volume $V_{BS}$ using recent progress in experimental Gaussian boson sampling and its classical simulation. Oh \textit{et al.}~proposed a tensor-network-based classical algorithm for simulating lossy Gaussian boson sampling and applied it to several large-scale experiments, including Jiuzhang-type platforms \cite{Oh24}. Their analysis emphasizes that photon loss can substantially reduce the computationally relevant quantum resource. More recently, Liu \textit{et al.}~reported a substantially larger Gaussian boson sampling experiment using 1,024 squeezed states in 8,176 modes \cite{Liu26}. These developments provide useful test cases for the PQD-based scalability bound and for the proposed metrics $S_{BS}$ and $V_{BS}$.

Using the experimental parameters reported for each dataset \cite{Zhong20,Zhong21,Liu26}, we estimate $M_B$, $S_{BS}$, and the volume $V_{BS}(M_{L})$, where we set $M_{L}$ to the implemented mode size. The results are summarized in Table \ref{table:Jiuzhang}.
A representative calculation showing how the experimentally reported quantities are mapped onto the parameters of the present framework is provided for Jiuzhang4.0 in Sec.~IV of the Supplementary Material.

\begin{table}[b]
 \caption{PQD-based scalability bound $M_B$, $S_{BS}$, and $V_{BS}$ for representative Jiuzhang-type Gaussian boson sampling experiments. P125-1 and P125-2 denote Jiuzhang2.0 datasets measured with a 125 $\mu$m pump waist at pump powers of 0.5 W and 1.412 W, respectively. The label 6 nJ specifies the pump pulse energy used for the selected Jiuzhang4.0 dataset.}
 \begin{center}
\begin{tabular}{cccccc}
\noalign{\smallskip} \noalign{\smallskip}
\hline\hline
\noalign{\smallskip}
\multicolumn{1}{c}{\,\,Experiment\,\,}
& Dataset    & \,\,\,\,\,\,$M$\,\,\,   \,\,\,& $M_B$\,\,\,              & \,\,\,\,\,$S_{BS}$\,\,\,\,\,\, & \,\,\,$V_{BS}$\,\,\, \\ 
\hline \noalign{\smallskip}
\makecell{Jiuzhang1.0 \\ \cite{Zhong20}}   & 100-mode  	& 100   & $8.96 \times 10^3$  & 69.17  & 10.00 \\ \noalign{\smallskip} \hline \noalign{\smallskip}
\makecell{Jiuzhang2.0 \\ \cite{Zhong21}}   & P125-1   & 144    & $5.96 \times 10^4$ & 20.42   & 12.00  \\ \noalign{\smallskip} \hline \noalign{\smallskip}
\makecell{Jiuzhang2.0 \\ \cite{Zhong21}}  & P125-2  & 144    & $6.16 \times 10^4$  & 60.20  & 12.00 \\ \noalign{\smallskip} \hline \noalign{\smallskip}
\makecell{Jiuzhang4.0 \\ \cite{Liu26}}      & 6 nJ     & 8176  & N/A  & 3793 & 90.42 \\ \noalign{\smallskip} \hline\hline
\end{tabular}
\end{center}
\label{table:Jiuzhang}
\end{table}

The first important observation is that the PQD-based classical simulatability bound does not limit the boson sampling score $S_{BS}$ for these datasets. For the Jiuzhang1.0 and Jiuzhang2.0 datasets, the estimated scalability bound $M_B$ is much larger than the implemented mode number. In the case of Jiuzhang4.0, $M_B$ is not uniquely determined since the mode-number dependence of the transmission is not specified for its hybrid spatial-temporal architecture. At the implemented scale, however, $\eta_L(8176) \gg \eta_{th}$, so the PQD-based classical simulatability condition is not satisfied and $S_{BS}$ remains nonzero (see the Supplementary Material for details). Therefore, the present PQD-based classical simulatability bound does not certify efficient classical simulation for any of these datasets.

For the 100-mode Jiuzhang1.0 dataset, $M_B$ is far above the implemented mode size $M=100$. However, the volume is limited by the low-collision condition. Thus, even though the score is 69.17, the effective boson sampling volume remains $V_{BS}(100)=\sqrt{100}=10$ limited by the available mode number. The same behavior is observed in the Jiuzhang2.0 datasets. For Jiuzhang2.0 P125-1, the boson sampling score gives $S_{BS}\simeq20.42$ and the volume is $V_{BS}(144)=12.00$. Jiuzhang2.0 P125-2 has a higher score $S_{BS}\simeq60.20$ than the P125-1 dataset. Nevertheless, its volume remains constrained by the mode-size limit, yielding $V_{BS}(144)=\sqrt{144}=12$. Therefore, for these earlier Jiuzhang-type experiments, increasing the number of input photons increases the score but does not increase the volume beyond the low-collision bound.

Jiuzhang4.0 shows a qualitatively larger scale. As discussed above, $\eta_L(8176) \gg \eta_{th}$ indicates that the PQD-based classical simulatability condition is not satisfied. The corresponding boson sampling score is $S_{BS} \simeq 3793$. This large score reflects the large number of detected photons together with the assumed high mode match. However, $V_{BS}$ is still limited by the low-collision condition, $V_{BS}(8176) \simeq 90.42$. Thus, within the present metric, Jiuzhang4.0 exceeds the $V_{BS}=50$ benchmark but remains below $V_{BS}=100$. The limitation is not the boson sampling score but the finite number of modes: reaching $V_{BS}=100$ requires at least $M\geq10^4$, even if the score is already larger than 100. 
Note that the reported quantum computational advantage is instead supported by empirical benchmarking against the best currently available classical methods, including loss-exploiting tensor-network simulations \cite{Oh24,Liu26}. 

Since all considered datasets in Table \ref{table:Jiuzhang} satisfy $M \ll M_B$ or $\eta_L \gg \eta_{th}$, the PQD-based classical simulatability condition is not the present bottleneck. Instead, once $S_{BS} \geq \sqrt{M}$, increasing the detected photon number alone no longer improves $V_{BS}$, and further scaling requires a larger number of optical modes together with maintaining high transmission and mode overlap. 

The proposed metric can also be used to estimate a low-collision operating point for the existing Jiuzhang4.0 interferometer. For $M=8176$, the low-collision boundary corresponds to a mean detected photon number of $\bar{N}_D \simeq \sqrt{8176} \simeq 90.42$. Assuming uniform squeezing across the 1,024 populated inputs and unchanged overall system efficiency and mode overlap, this condition gives $\mu=\bar{N}_D/(1024\times0.51)\simeq0.173$, corresponding to an effective squeezing parameter $r=\operatorname{arcsinh}\sqrt{\mu}\simeq0.40$. At this squeezing level, $\eta_{th}\simeq2.19\times10^{-5}$ remains much smaller than $\eta_L(8176)$, so the PQD-based classical-simulability condition is still not satisfied. Thus, reducing the pump pulse energy could bring the existing 8176-mode interferometer close to the low-collision boundary without crossing the PQD-based simulability bound. 
\section{Conclusion}
In conclusion, we develop a framework for assessing the scalability of boson sampling devices under realistic experimental constraints using the non-negativity conditions of the PQD functions. By incorporating photon loss, detector efficiency, dark counts, and mode mismatch, we derive an architecture-dependent scalability bound $M_B$. 
Based on this bound, we introduce the boson sampling score $S_{BS}$ and the boson sampling volume $V_{BS}$. The boson sampling score quantifies the usable capacity of a device by estimating an effective number of detected, indistinguishable photons after experimental constraints are included, whereas the volume maps this capacity onto an equivalent ideal low-collision boson sampling size. Consequently, $V_{BS}$ provides a practical benchmark for assessing access to an advantage-relevant problem size regime.

Applying the proposed framework to architectures and recent Gaussian boson sampling experiments identifies which factor limits device performance: the PQD-based classical simulatability bound $M_B$, the boson sampling score $S_{BS}$, or the available number of optical modes. The architecture comparison further shows how differences in interferometer depth scaling affect the achievable $S_{BS}$ and $V_{BS}$. A larger $S_{BS}$ indicates a greater effective number of detected photons available for quantum interference, whereas a larger $V_{BS}$ indicates access to a larger effective low-collision problem size. Both are favorable indicators of device performance. The analysis of the Jiuzhang datasets considered here further shows that increasing the detected photon number alone does not increase $V_{BS}$ once the available number of modes becomes the dominant constraint. Thus, increasing $S_{BS}$ must be accompanied by sufficient mode capacity to yield a larger $V_{BS}$. Further scaling therefore requires a larger number of optical modes while maintaining high transmission and mode overlap. The proposed metrics provide a compact link between experimentally accessible device parameters and the computationally relevant scale of boson sampling, offering practical guidance for the comparison and design of linear-optical boson sampling experiments.

\section*{Data availability}
There are no publicly available research data or software supporting this manuscript. Requests for further information or data should be sent to the authors.

\section*{Funding}
This work was supported in part by the National Research Foundation of Korea (RS-2023-00208500, RS-2024-00442762, RS-2023-NR068116, RS-2025-03532992), the Institute of Information \& Communications Technology Planning \& Evaluation (IITP) grant funded by the Korea government (MSIT) (RS-2025-02263264, No.~2019-0-00003), and the Information Technology Research Center support program (IITP-2025-RS-2022-00164799). J.H.~also acknowledges support from the Yonsei University Research Fund (Project No. 2025-22-0140).


\end{document}